\documentclass[11pt]{article}

\usepackage[a4paper,margin=1in]{geometry}
\usepackage{amsmath,amssymb,amsthm,mathtools}
\usepackage{booktabs}
\usepackage{xcolor}
\usepackage[colorlinks=true,citecolor=blue,linkcolor=blue,urlcolor=blue]{hyperref}
\usepackage[
  backend=biber,
  style=authoryear,
  natbib=true,
  maxcitenames=2,
  maxbibnames=99,
  giveninits=true,
  doi=true,
  url=false,
  uniquename=init
]{biblatex}
\AtEveryCite{\color{blue}}

\newtheorem{fact}{Fact}

\newcommand{\R}{\mathcal R}
\newcommand{\Htrue}{\mathcal H_0}

\title{Where Does the Union Bound Go?\\
Best-Arm Identification and Strong FWER Control}
\author{Rianne de Heide\\Department of Applied Mathematics, University of Twente\\ and Centrum Wiskunde \& Informatica, Amsterdam\\\texttt{r.deheide@utwente.nl}}
\date{20 August 2026}

\begin{document}
\maketitle

\begin{abstract}
In fixed-confidence best-arm identification, proofs often use a union bound
across the competing arms. From a multiple-testing point of view this can look
puzzling: if the best arm is unique, only one hypothesis of the form ``arm
$i$ is best'' can be true. Why then should there be a Bonferroni-type factor
of $K-1$? The answer is that there are two natural ways to orient the
hypotheses. In one orientation, best-arm identification is literally
a strong familywise-error-rate (FWER) problem with $K-1$ true nulls. In the opposite orientation, exactly one
null is true, but a pairwise implementation can falsely reject that one null
through any of $K-1$ comparisons. Thus the multiplicity has not disappeared;
it just pops up in different places. This note makes the equivalence explicit in the terminology of both communities.
\end{abstract}

\section{The puzzle}

Suppose there are $K$ arms with means
\[
  \mu_1,\ldots,\mu_K.
\]
An ``arm'' is simply one candidate option (for example, a treatment or
system) from which the algorithm can collect observations.
Let
\[
  \Theta_{\mathrm{uniq}}
  :=\bigl\{\mu:\lvert\arg\max_{i\in\{1,\ldots,K\}}\mu_i\rvert=1\bigr\}
\]
be the unique-best-arm parameter space, and assume throughout that
$\mu\in\Theta_{\mathrm{uniq}}$. Then the unique best arm is
\[
  i^\star=\arg\max_{i\in\{1,\ldots,K\}}\mu_i.
\]
In fixed-confidence best-arm identification (BAI), an adaptive algorithm
samples the arms, stops at a data-dependent time $\tau$, and returns an arm
$\widehat i_\tau$. We use $\delta$-correctness below for the error-control
requirement
\begin{equation}
  \Pr_\mu(\widehat i_\tau\neq i^\star)\leq\delta
  \qquad\text{for every instance with a unique best arm.}
  \label{eq:BAIcorrect}
\end{equation}
In the standard fixed-confidence BAI definition, a $\delta$-PAC strategy
is additionally required to stop almost surely, i.e.
$\Pr_\mu(\tau<\infty)=1$; PAC stands for ``probably approximately correct''.
See, for example, \textcite{GarivierKaufmann2016}.

A standard BAI proof often reaches an event of the form
\[
  \{\widehat i_\tau\neq i^\star\}
  \subseteq
  \bigcup_{j\neq i^\star}\mathcal E_j
\]
and then uses
\[
  \Pr_\mu(\widehat i_\tau\neq i^\star)
  \leq
  \sum_{j\neq i^\star}\Pr_\mu(\mathcal E_j).
\]
This is simply the union bound, and it commonly leads to an error allocation such as $\delta/(K-1)$ across the competing events;
in logarithmic stopping thresholds this typically appears as an additive
$\log(K-1)$ contribution. Modern analyses can make the underlying time-uniform concentration
much sharper, but a union over the $K-1$ wrong answers still appears in
standard BAI calibrations \parencite{KaufmannKoolen2021}.

Now compare this with a familiar multiple-testing observation. If a family of
null hypotheses is logically constrained so that at most $m_0$ of them can be
true simultaneously, a Bonferroni proof of strong familywise-error control
only needs to sum over those $m_0$ true nulls. But among the statements
\[
  \text{``arm 1 is best''},\ldots,\text{``arm $K$ is best''},
\]
exactly one is true. So why is the BAI proof paying for $K-1$ events?

The apparent contradiction comes from implicitly switching between
two different families of null hypotheses.

\section{Multiple testing: strong FWER control}

For a family of null hypotheses $H_1,\ldots,H_m$, let $\R$ be the set of
rejected hypotheses and define the index set of true nulls by
$\Htrue(\mu):=\{i:\mu\in H_i\}$. Strong familywise-error-rate (FWER) control at level $\alpha$
means
\begin{equation}
  \Pr_\mu\bigl(\R\cap\Htrue(\mu)\neq\varnothing\bigr)\leq\alpha
  \qquad\text{for every }\mu.
  \label{eq:FWER}
\end{equation}
In words: the probability of rejecting at least one true null is at most
$\alpha$, whatever the configuration of true and false nulls.

The elementary Bonferroni argument is
\[
  \Pr_\mu\bigl(\R\cap\Htrue(\mu)\neq\varnothing\bigr)
  \leq
  \sum_{i\in\Htrue(\mu)}
  \Pr_\mu(H_i\text{ is rejected}).
\]
The sum is over the \emph{true} nulls, not automatically over every hypothesis
in the family. Logical information about which nulls can be true together can
therefore matter substantially; this is one of the motivations behind the
partitioning principle and related structured multiple-testing methods
\parencite{FinnerStrassburger2002}.

The same basic idea---use the logical restrictions among the hypotheses
rather than pretending that every truth configuration is possible---also
underlies classical improvements of Bonferroni such as Shaffer's procedure and
is closely related to closed testing
\parencite{Shaffer1986,MarcusPeritzGabriel1976}.

\begin{center}
\fcolorbox{black}{white}{%
\begin{minipage}{0.90\linewidth}
\color{black}
\textbf{The partitioning principle\\}
Suppose the parameter space is split into disjoint regions
$P_1,\ldots,P_M$ that cover all possibilities. Exactly one region is true.
If every $P_m$ is tested at level $\alpha$, there is no need to divide
$\alpha$ by $M$: the event ``reject at least one true partition hypothesis''
is simply the event ``reject the unique true region'', which has probability at
most $\alpha$.

Now suppose an original null hypothesis $H$ is the union of several of these
regions. Reject $H$ only when \emph{all} partition regions contained in $H$ have
been rejected. If $H$ is true, the unique true cell lies inside $H$, so
rejecting $H$ necessarily entails rejecting that true cell. Hence the induced
decisions on the original hypotheses strongly control FWER at level
$\alpha$. This is the partitioning principle: first encode the logical
structure into mutually exclusive cases, and only then test
\parencite{FinnerStrassburger2002}.
\end{minipage}%
}
\end{center}

\section{Orientation 1: ``arm \texorpdfstring{$i$}{i} is not best''}

Define
\begin{equation}
  H_i^-:=\{\mu\in\Theta_{\mathrm{uniq}}: i\neq i^\star(\mu)\}.
  \label{eq:notbestnull}
\end{equation}
The null $H_i^-$ says that arm $i$ is \emph{not} the best arm. If a BAI
procedure recommends arm $i$, interpret this as rejecting $H_i^-$. Since the
procedure returns one arm, its rejection set is simply
\[
  \R=\{\widehat i_\tau\}.
\]

Suppose the true best arm is $i^\star$. Then
\[
  H_{i^\star}^-\text{ is false},
  \qquad
  H_i^-\text{ is true for every }i\neq i^\star.
\]
So there are exactly $K-1$ true nulls.

\begin{fact}
For the family \eqref{eq:notbestnull}, the BAI error event is exactly the FWER
event:
\[
  \{\widehat i_\tau\neq i^\star\}
  =
  \{\R\cap\Htrue(\mu)\neq\varnothing\}.
\]
Consequently, $\delta$-correct BAI is strong FWER control at level $\delta$ for
this logically constrained family over the parameter space
$\Theta_{\mathrm{uniq}}$.
\end{fact}

In this orientation there is therefore no puzzle at all. The multiple-testing
rule ``only pay for hypotheses that can be true at the same time'' says that a naive
Bonferroni proof may have to pay for $K-1$ hypotheses, because $K-1$ of these
nulls \emph{are} simultaneously true.

\section{Orientation 2: ``arm \texorpdfstring{$i$}{i} is best''}

Now reverse the hypotheses. Define
\begin{equation}
  G_i:=\{\mu\in\Theta_{\mathrm{uniq}}:i=i^\star(\mu)\}.
  \label{eq:bestnull}
\end{equation}
The $G_i$ form a partition of the unique-best-arm parameter space. Exactly one
of them is true. Thus the $G_i$ are literally partition hypotheses in
the sense of the box in Section~2.

Imagine an elimination algorithm: reject $G_i$ when there is sufficient
evidence that arm $i$ cannot be best, eliminate arm $i$, and stop when only
one arm remains. In this formulation, the multiple-testing intuition is
correct in a very literal way.

\begin{fact}
If each $G_i$ is tested by an anytime-valid level-$\delta$ test, meaning
that its rejection event satisfies
\[
  \sup_{\mu\in G_i}
  \Pr_\mu\!\left(\exists t:\ G_i\text{ is rejected by time }t\right)
  \leq\delta,
\]
with respect to the filtration generated by the adaptively sampled bandit
experiment, then an elimination procedure that returns the last non-rejected
cell is $\delta$-correct. No multiplicity correction across
$G_1,\ldots,G_K$ is needed.
\end{fact}

So where did the $K-1$ go? It appears when we ask how a test of the \emph{one}
true composite hypothesis $G_{i^\star}$ is constructed.

Equivalently,
\begin{equation}
	G_i
	=
	\bigcap_{j\neq i}\{\mu_i>\mu_j\}.
	\label{eq:intersection}
\end{equation}
Thus ``arm $i$ is best'' is an intersection of $K-1$ pairwise statements. A
pairwise elimination rule rejects $G_i$ as soon as one competitor has produced
sufficient evidence against it. If $R_{j,i}$ denotes the event that the
pairwise comparison with $j$ eliminates $i$, then
\[
  \{G_i\text{ is rejected}\}
  =
  \bigcup_{j\neq i}R_{j,i}
\]
for such a construction. Under the true best-arm null $G_{i^\star}$, all
$K-1$ pairwise inequalities
\[
  \mu_{i^\star}>\mu_j,\qquad j\neq i^\star,
\]
are true. Hence
\begin{equation}
  \Pr_\mu(G_{i^\star}\text{ is rejected})
  \leq
  \sum_{j\neq i^\star}\Pr_\mu(R_{j,i^\star}).
  \label{eq:insideunion}
\end{equation}
The Bonferroni correction has moved inside the test of the single true
composite null.

\section{What this says and what not, and related work}

The point of the observation is logical, not that every union bound used in
BAI is necessary or sharp. Once we recognize that ``arm $i$ is best'' is one
composite null, it is natural to ask whether it can be tested directly rather
than by $K-1$ separately calibrated pairwise tests. Two classical
statistical traditions are especially close to this question: ranking and
selection, and multiple comparisons with the best (MCB). Bechhofer's early
ranking-and-selection formulation explicitly recast ranking normal means as a
multiple-decision problem \parencite{Bechhofer1954}. Paulson subsequently gave
a sequential procedure for selecting the population with the largest mean in
which inferior populations can be eliminated as sampling proceeds
\parencite{Paulson1964}. The MCB literature studies comparisons of each
treatment with its competitors and includes both sequential and simultaneous
procedures \parencite{HsuEdwards1983,EdwardsHsu1983,Hsu1984}. Hsu's stepwise
MCB procedure is particularly close in spirit to an elimination viewpoint: it
successively identifies treatments as strictly not the best
\parencite{Hsu1992}. These procedures share the same geometry as BAI:
statements about a treatment being best are built from comparisons with its
competitors. On the multiple-testing side, the partitioning and closed-testing
viewpoints make explicit that logical relations among hypotheses can reduce
multiplicity, while Shaffer-type procedures exploit the restricted sets of
hypotheses that can be true simultaneously
\parencite{MarcusPeritzGabriel1976,Shaffer1986,FinnerStrassburger2002}.

The factor $K-1$ can be seen particularly cleanly in the development of
fixed-confidence BAI. In the Bernoulli case, Theorem~10 of
\textcite{GarivierKaufmann2016} proves $\delta$-correctness of Chernoff's
stopping rule with threshold
\[
  \beta(t,\delta)=\log\!\left(\frac{2t(K-1)}{\delta}\right).
\]
The proof first fixes the true best arm and then takes a union over the
$K-1$ possible wrong winners. The $K-1$ there counts the $K-1$ distinct ways
in which the one true best-arm statement can be contradicted by a competitor.

\textcite{KaufmannKoolen2021} sharpen this analysis substantially.
Their mixture-martingale inequalities are time-uniform and can combine
evidence from several arms, avoiding a cruder arm-by-arm concentration
argument.
Nevertheless, in their Section~5.1 the BAI error event, with arm $1$ taken to
be the true best arm, is again bounded by an event of the form
\[
  \{\exists t,\ \exists a\neq 1:\ \text{the pair }\{1,a\}
    \text{ crosses its boundary}\}.
\]
They then explicitly apply a weighted union bound over the $K-1$ subsets
$\{1,2\},\ldots,\{1,K\}$. More precisely, their resulting threshold is
\[
  \widehat c_t(\delta)
  =6\log\!\left(\log\!\left(\frac{t}{2}\right)+1\right)
   +2C_{\exp}\!\left(\frac{1}{2}\log\!\left(\frac{K-1}{\delta}\right)\right),
\]
where $C_{\exp}(x)\sim x$ as $x\to\infty$. Thus the multiplicity enters through
$\log((K-1)/\delta)$ as the argument of the calibration function. Its dependence on time is much sharper
than the earlier $\log(2t(K-1)/\delta)$ boundary. This is exactly the
distinction highlighted in Section~4: the concentration within a pair can be
sharpened without making the $K-1$ possible pairwise witnesses disappear.

The same paper introduces the \emph{rank} of a pure-exploration
identification problem (that is, a sequential learning problem whose aim is to
identify a correct answer rather than to maximize cumulative reward). BAI has
rank $2$, because to show that a proposed arm
$i$ is not best it suffices to exhibit one competitor $a$ with
$\mu_a>\mu_i$; each individual witness therefore involves only two arms.
But there may still be $K-1$ such witnesses. The rank and the number of
possible witnesses are consequently different notions: rank $2$ explains why
pairwise statistics suffice, while the union over the possible competitors
explains why $K-1$ can remain in a nonasymptotic error calibration
\parencite{KaufmannKoolen2021}.

This also clarifies what should \emph{not} be concluded. These proofs do not say that every BAI method must incur
a multiplicative $K-1$ loss in sample complexity. For fixed $K$ and
$\delta\downarrow0$, the classical BAI lower bound and the asymptotically
optimal Track-and-Stop analysis are governed at leading order by
$\log(1/\delta)$; inserting $K-1$ inside the logarithm changes only a
lower-order term \parencite{GarivierKaufmann2016}. Also, Bonferroni is not the only
possible way, as the vast multiple testing literature shows, to construct a
test of $G_i$: a joint test can exploit dependence and the geometry of the
full composite null. There are concrete examples of this possibility in
related pure-exploration problems: \textcite{KatzSamuelsEtAl2020} use
empirical-process tools and the Gaussian width of the arm set to obtain linear
and combinatorial bandit guarantees that avoid an explicit union bound over the
number of arms. This is not the same as removing the $K-1$ term from ordinary
unstructured BAI, but it illustrates that joint geometric control can replace
an arm-by-arm union bound in a closely related identification problem.

Finally, there are nearby BAI formulations that make the multiple testing
language even more explicit. A representative confidence-bound method is
LUCB, whose stopping logic compares lower and upper confidence bounds
\parencite{KalyanakrishnanEtAl2012}, while \textcite{GarivierKaufmann2021}
study sequential identification directly as testing among possibly
overlapping hypotheses. So the similarities are plenty:
both communities are organizing error across logically related candidate
answers, but they often place the multiplicity at different levels of the
construction.

\printbibliography

@inproceedings{GarivierKaufmann2016,
  author    = {Garivier, Aur{\'e}lien and Kaufmann, Emilie},
  title     = {Optimal Best Arm Identification with Fixed Confidence},
  booktitle = {Proceedings of the 29th Conference on Learning Theory},
  series    = {Proceedings of Machine Learning Research},
  volume    = {49},
  pages     = {998--1027},
  year      = {2016},
  publisher = {PMLR}
}

@article{KaufmannKoolen2021,
  author       = {Kaufmann, Emilie and Koolen, Wouter M.},
  title        = {Mixture Martingales Revisited with Applications to Sequential Tests and Confidence Intervals},
  journaltitle = {Journal of Machine Learning Research},
  year         = {2021},
  volume       = {22},
  number       = {246},
  pages        = {1--44}
}

@article{FinnerStrassburger2002,
  author       = {Finner, Helmut and Strassburger, Klaus},
  title        = {The Partitioning Principle: A Powerful Tool in Multiple Decision Theory},
  journaltitle = {The Annals of Statistics},
  year         = {2002},
  volume       = {30},
  number       = {4},
  pages        = {1194--1213},
  doi          = {10.1214/aos/1031689023}
}

@article{HsuEdwards1983,
  author       = {Hsu, Jason C. and Edwards, David G.},
  title        = {Sequential Multiple Comparisons with the Best},
  journaltitle = {Journal of the American Statistical Association},
  year         = {1983},
  volume       = {78},
  number       = {384},
  pages        = {958--964},
  doi          = {10.1080/01621459.1983.10477046}
}

@article{EdwardsHsu1983,
  author       = {Edwards, Donald G. and Hsu, Jason C.},
  title        = {Multiple Comparisons with the Best Treatment},
  journaltitle = {Journal of the American Statistical Association},
  year         = {1983},
  volume       = {78},
  number       = {384},
  pages        = {965--971},
  doi          = {10.1080/01621459.1983.10477047}
}

@article{Hsu1984,
  author       = {Hsu, Jason C.},
  title        = {Constrained Simultaneous Confidence Intervals for Multiple Comparisons with the Best},
  journaltitle = {The Annals of Statistics},
  year         = {1984},
  volume       = {12},
  number       = {3},
  pages        = {1136--1144},
  doi          = {10.1214/aos/1176346732}
}

@article{Hsu1992,
  author       = {Hsu, Jason C.},
  title        = {Stepwise Multiple Comparisons with the Best},
  journaltitle = {Journal of Statistical Planning and Inference},
  year         = {1992},
  volume       = {33},
  number       = {2},
  pages        = {197--204},
  doi          = {10.1016/0378-3758(92)90067-3}
}

@article{Bechhofer1954,
  author       = {Bechhofer, Robert E.},
  title        = {A Single-Sample Multiple Decision Procedure for Ranking Means of Normal Populations with Known Variances},
  journaltitle = {The Annals of Mathematical Statistics},
  year         = {1954},
  volume       = {25},
  number       = {1},
  pages        = {16--39},
  doi          = {10.1214/aoms/1177728845}
}

@article{Paulson1964,
  author       = {Paulson, Edward},
  title        = {A Sequential Procedure for Selecting the Population with the Largest Mean from {$k$} Normal Populations},
  journaltitle = {The Annals of Mathematical Statistics},
  year         = {1964},
  volume       = {35},
  number       = {1},
  pages        = {174--180},
  doi          = {10.1214/aoms/1177703739}
}

@article{MarcusPeritzGabriel1976,
  author       = {Marcus, Ruth and Peritz, Eric and Gabriel, K. R.},
  title        = {On Closed Testing Procedures with Special Reference to Ordered Analysis of Variance},
  journaltitle = {Biometrika},
  year         = {1976},
  volume       = {63},
  number       = {3},
  pages        = {655--660},
  doi          = {10.1093/biomet/63.3.655}
}

@article{Shaffer1986,
  author       = {Shaffer, Juliet Popper},
  title        = {Modified Sequentially Rejective Multiple Test Procedures},
  journaltitle = {Journal of the American Statistical Association},
  year         = {1986},
  volume       = {81},
  number       = {395},
  pages        = {826--831},
  doi          = {10.1080/01621459.1986.10478341}
}

@inproceedings{KalyanakrishnanEtAl2012,
  author    = {Kalyanakrishnan, Shivaram and Tewari, Ambuj and Auer, Peter and Stone, Peter},
  title     = {{PAC} Subset Selection in Stochastic Multi-Armed Bandits},
  booktitle = {Proceedings of the 29th International Conference on Machine Learning},
  pages     = {655--662},
  year      = {2012},
  publisher = {Omnipress}
}

@article{GarivierKaufmann2021,
  author       = {Garivier, Aur{\'e}lien and Kaufmann, Emilie},
  title        = {Nonasymptotic Sequential Tests for Overlapping Hypotheses Applied to Near-Optimal Arm Identification in Bandit Models},
  journaltitle = {Sequential Analysis},
  year         = {2021},
  volume       = {40},
  number       = {1},
  pages        = {61--96},
  doi          = {10.1080/07474946.2021.1847965}
}

@inproceedings{KatzSamuelsEtAl2020,
  author    = {Katz-Samuels, Julian and Jain, Lalit and Karnin, Zohar and Jamieson, Kevin G.},
  title     = {An Empirical Process Approach to the Union Bound: Practical Algorithms for Combinatorial and Linear Bandits},
  booktitle = {Advances in Neural Information Processing Systems},
  volume    = {33},
  pages     = {10371--10382},
  year      = {2020}
}
\end{document}